\documentstyle[prl,aps,graphicx]{revtex}
\newcounter{one}
\newcounter{two}
\newcounter{three}
\newcounter{four}
\newcounter{five}
\begin{document}
\draft{}
\twocolumn[\hsize\textwidth\columnwidth\hsize
           \csname @twocolumnfalse\endcsname
%
\title{Oxygen isotope effect on the in-plane penetration depth in underdoped
La$_{2-\mbox{\footnotesize{x}}}$Sr$_{\mbox{\footnotesize{x}}}$CuO$_{4}$
single crystals}
\author{J.~Hofer$^{1}$,
K.~Conder$^{2}$,
T.~Sasagawa$^{3}$,
Guo-meng Zhao$^{1}$,
M.~Willemin$^{1}$,
H.~Keller$^{1}$,
and K.~Kishio$^{3}$}
\address{
$^{1}$Physik-Institut der Universit\"at Z\"urich, Winterthurerstr.
190, CH-8057 Z\"urich, Switzerland \\
$^{2}$Laboratorium f\"ur Festk\"orperphysik, ETH H\"{o}nggerberg Z\"urich,
CH-8093 Z\"urich, Switzerland \\
$^{3}$Department of Superconductivity, University of Tokyo, 7-3-1
Hongo, Bunkyo-ku, Tokyo 113-8656, Japan \\
}
\date{submitted to Phys. Rev. Lett., December 18 1999}
\maketitle
\begin{abstract}
We report measurements of the oxygen isotope effect (OIE) on the in-plane
penetration depth
$\lambda_{ab}(0)$ in underdoped
La$_{2-\mbox{\footnotesize{x}}}$Sr$_{\mbox{\footnotesize{x}}}$CuO$_{4}$
single crystals.
A highly sensitive magnetic
torque sensor with a resolution of $\Delta\tau \simeq 10^{-12}$ Nm was
used for the magnetic measurements on microcrystals with a mass
of $\approx 10 \ \mu$g.
The OIE on
$\lambda_{ab}^{-2}(0)$ is found to be -10(2)\% for $x = 0.080$ and
-8(1)\% for $x = 0.086$. It arises mainly from the oxygen mass
dependence of the in-plane effective mass $m_{ab}^{\ast}$.
The present results suggest that lattice vibrations are important for the
occurrence of high temperature superconductivity.
\end{abstract}
\pacs{PACS numbers: 74.25.Ha, 74.20.Mn, 82.20.Tr}
%
]
%
Soon after the discovery of high temperature superconductivity
\cite{BednorzZPBCM86} a large number of isotope effect
experiments were performed to
investigate the pairing mechanism \cite{FranckBOOK94}. The very
first $^{16}$O/$^{18}$O
isotope studies were carried out on optimally doped samples and
showed a negligible oxygen isotope effect (OIE) \cite{BatloggPRL87}.
A number of subsequent experiments revealed a
dependence of $T_{c}$ on the oxygen isotope mass
$M_{O}$ \cite{BatloggPRL287,CrawfordPRB90,ZechNAT94} and
on the copper isotope mass $M_{Cu}$ \cite{FranckPRL93,ZhaoPRB96}. It was
generally found that the isotope effects are large in the
underdoped region, but become small when the doping increases towards
the optimally-doped and overdoped regimes \cite{CrawfordPRB90,ZhaoPRB96}.
A large OIE on the Meissner
fraction was observed in
La$_{2-\mbox{\footnotesize{x}}}$Sr$_{\mbox{\footnotesize{x}}}$CuO$_{4}$
powder samples and attributed to a strong oxygen mass dependence of the
effective mass
$m^{\ast}$ of the superconducting charge carriers \cite{ZhaoNAT97}.
However, these experiments were done on powder samples and thus probed the
average magnetic properties of this highly anisotropic superconductor.
For a quantitative analysis isotope experiments on single crystals are
required.

Unfortunately, a complete oxygen isotope exchange by diffusion is
very difficult in single crystals with a large volume, as shown by a
study on
Bi$_{2}$Sr$_{2}$CaCu$_{2}$O$_{8 + \delta}$ crystals with
$V \approx 5 \times 4 \times 0.1$ mm$^{3}$ \cite{MartinPC95}.
Indeed, our preliminary investigations on
La$_{2-\mbox{\footnotesize{x}}}$Sr$_{\mbox{\footnotesize{x}}}$CuO$_{4}$
single crystals with $V \simeq 1 \times 1 \times
0.3$ mm$^{3}$ showed that a complete isotope exchange was not possible.
In order to reach a complete oxygen-isotope exchange, microcrystals
with a
volume of only $V \approx 150 \times 150 \times 50 \ \mu$m$^{3}$
(mass $\approx 10 \ \mu$g) were
used for the present study. In these tiny samples,having a volume not very
much larger than the grain size of
polycrystalline samples, an almost complete oxygen isotope exchange
was achieved by diffusion, as shown below.

It is known that the transition temperature $T_{c}$ and the in-plane
penetration depth
$\lambda_{ab}$
of a cuprate superconductor can be
determined from temperature- and field-dependent measurements of the
reversible magnetization $M$ using SQUID magnetometry \cite{LiPRB93}. Close
to $T_{c}$ the magnetic moment $m = V M$ of microcrystals with a
mass of $\approx 10 \ \mu$g lies well below the resolution
$\Delta m = 10^{-10}$ Am$^{2}$ of commercial SQUID magnetometers.
Therefore, all magnetic measurements were carried out using a
highly sensitive torque magnetometer with a resolution
$\Delta \tau < 10^{-12}$ Nm \cite{WilleminJAP98}. The magnetic torque
$\vec{\tau} = \vec{m} \times \vec{B_{a}}$ is usually recorded as a
function of the angle $\delta$ between the field $\vec{B_{a}}$ and the $c$
axis of the crystal \cite{FarrellPRL88,HoferPC98}. However, when
$\delta$ is fixed at a finite value, temperature- and field-dependent
torque measurements can be performed a well. An appropriate
angle to carry out these measurements is $\delta = 45^{\circ}$ for the
following reasons:
(\roman{one}) $\vec{m}$ is still pointing along the $c$ axis due to
the large anisotropy \cite{KoganPRB88}. (\roman{two}) The
magnetic torque $\tau = M B_{a} \sin(\delta)$ is sufficiently large to be
measured for tiny magnetic moments in small fields. (\roman{three}) The
reversible regime in the ($B_{a},T$) phase diagram is almost as large as for
$\delta = 0^{\circ}$ \cite{BlatterRMP94}, and a thermodynamic
analysis of the measurements is possible over a wide temperature range.
Thus, torque measurements performed at fixed $\delta = 45^{\circ}$ can
be used to determine $T_{c}$ from the temperature-dependent
magnetization $M \propto \tau$, and to extract $\lambda_{ab}$ from
the field-dependent magnetization $M \propto \tau/B_{a}$.

Four microcrystals were cut
from single crystals  with Sr contents $x = 0.080$ (samples
\Roman{one}$a$ and \Roman{one}$b$) and $x = 0.086$ (samples
\Roman{two}$a$ and \Roman{two}$b$),
grown by the traveling-solvent-floating-zone method
\cite{SasagawaPRL98}. Underdoped samples were chosen for this study because
the
OIE is expected to be large in this doping regime \cite{CrawfordPRB90}.
For both sets of samples, \Roman{one} ($x = 0.080$) and
\Roman{two} ($x = 0.086$), the oxygen exchange procedure was as
follows:
First, both samples $a$ and $b$
were annealed in $^{16}$O in order to saturate the oxygen content.
Then sample $a$ was exchanged to $^{18}$O in an atmosphere with
97\% $^{18}$O while sample $b$ was simultaneously treated in $^{16}$O.
Finally, sample $a$ was back-exchanged to
$^{16}$O while
sample $b$ was exchanged to $^{18}$O. All exchange procedures were
performed in 1 bar atmosphere at 950 $^{\circ}$C for 50 h. The samples
were cooled to room temperature
\begin{figure}[htb]
        \centering
        \includegraphics[width=0.8\linewidth]{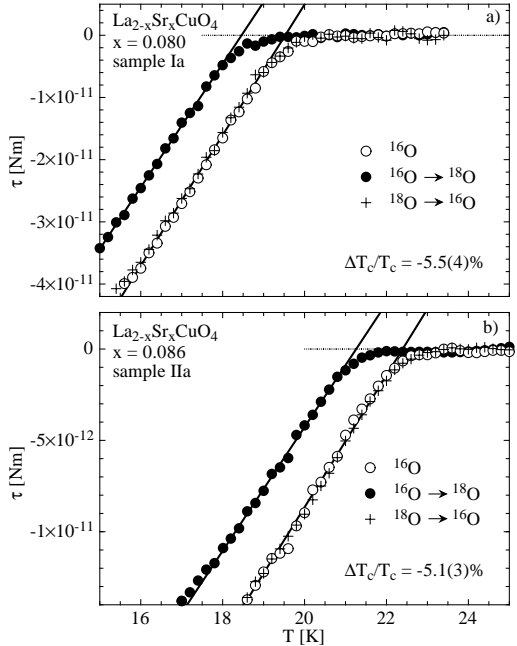}
        \caption[~]{Magnetic torque $\tau$ versus temperature, showing
        the OIE on $T_{c}$ for samples
        \Roman{one}$a$ and \Roman{two}$a$.
        The reproducibility of the exchange procedure,
        as checked by the back-exchange (cross symbols), demonstrates a
        complete isotope exchange.
        For clarity not all measured data points are shown.}
	\protect\label{Tc-shift}
\end{figure}
\noindent
with a cooling rate of 25 $^{\circ}$C/h.

In order to measure the magnetic torque, the samples were mounted on a
miniaturized
cantilever with piezoresistive readout and integrated calibration loop
\cite{WilleminJAP98}.
The cantilever was placed between the poles of a conventional NMR magnet
with a
maximal field $B_{a} = 1.5$ T.
The sensor was used in the so-called torsion mode
\cite{WilleminJAP98}, where major background effects arising from the strong
temperature and magnetic field dependence of the piezoresistive paths are
canceled out. In fact, the remaining
temperature-dependent background of the cantilever was sufficiently small
to perform temperature-dependent magnetic torque measurements.

The superconducting transition was studied by cooling the sample in a magnetic
field $B_{a} = 0.1$ T applied at $\delta = 45^{\circ}$. The torque signal
was continuously recorded upon cooling the crystal at a
cooling rate of 0.01 K/s. In order to determine the
background signal of the cantilever, the measurement was repeated in
zero field and the data were subtracted from those of the
field cooled measurement. The magnetic torque versus temperature obtained for
the samples \Roman{one}$a$ and
\Roman{two}$a$ is shown in Fig.~\ref{Tc-shift}.
Clearly, $T_{c}$ is lower for the $^{18}$O exchanged samples. We define
$T_{c}$ as the temperature where the linearly extrapolated
transition slope intersects the base line ($\tau = 0$ Nm).
The relative changes
in $T_{c}$ are found to be $\Delta T_{c} / T_{c} =
[T_{c}(^{18}\mbox{O}) -
T_{c}(^{16}\mbox{O})] / T_{c}(^{16}\mbox{O}) = -5.5(4)\%$ for sample
\Roman{one}$a$ and $\Delta T_{c} /
T_{c} = -5.1(3)\%$ for sample
\Roman{two}$a$.
The samples \Roman{one}$b$ and \Roman{two}$b$ showed no
change in the superconducting transition after the second annealing
in $^{16}$O, which indicates a complete saturation of oxygen
during the first annealing procedure. The oxygen isotope shifts of
$T_{c}$ are summarized in Table~\ref{Tab1}. As expected they are larger
for the samples \Roman{one}$a$ and \Roman{one}$b$ with a smaller $x$
\cite{CrawfordPRB90,ZhaoJPCM98}.
As shown in Fig.~\ref{Tc-shift}, the magnetic signals of the back-exchanged
samples (cross symbols) coincide with those of the
$^{16}$O annealed samples (open circles). This result implies that a complete
back-exchange from the $^{18}$O to $^{16}$O isotope was achieved.
This is only possible if after the back-exchange procedure the $^{16}$O
enrichment
in the sample corresponds to the $^{16}$O concentration of the gas,
which is 100\% (the contamination of the $^{16}$O atmosphere by the
$^{18}$O isotope
removed from the crystal is less than 10ppm and thus negligible).
For the same reason, after exchanging $^{16}$O by $^{18}$O, the $^{18}$O
concentration of the sample should be the same as that of the exchange
atmosphere
(i.e. 97\% $^{18}$O).
The fact that the shift in $T_{c}$ is parallel,
with no broadening of the transition,
also demonstrates an almost complete isotope exchange.
The exponent $\alpha_{O}$ of the OIE on $T_{c}$ is defined by
$T_{c} \propto M_{O}^{\alpha_{O}}$.
Taking into account a 97\% exchange, we find
$\alpha_{O} = -(\Delta T_{c} / T_{c}) /
(\Delta M_{O} / M_{O})
= 0.47(2)$ for $x = 0.080$ and $\alpha_{O} = 0.40(2)$
for $x = 0.086$, which is in good agreement
with the results obtained for powder samples with similar doping
\cite{CrawfordPRB90,ZhaoJPCM98}.

\begin{figure}[htb]
        \centering
        \includegraphics[width=0.8\linewidth]{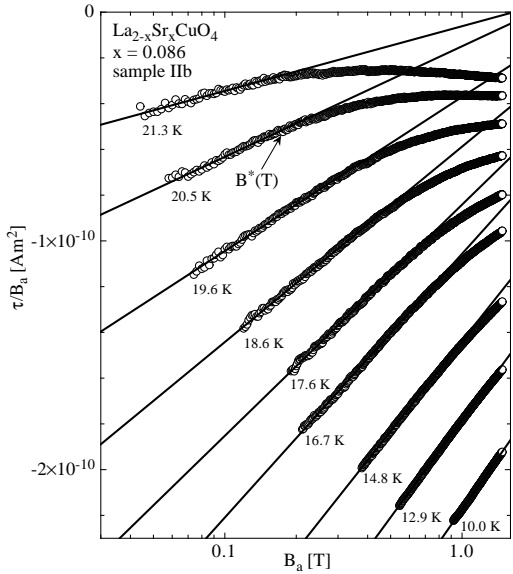}
        \caption[~]{Reversible part of the field-dependent torque
        $\tau/B_{a} \propto M$ versus $B_{a}$ for sample \Roman{two}$b$
        (after the second annealing in $^{16}$O). The measurements
        were performed at different temperatures at fixed $\delta =
        45^{\circ}$.
        $\lambda_{ab}^{-2}(T)$ is extracted from the slope of the linear
        part of the data for $B_{a} \leq B^{\ast}(T)$ (solid lines)
        by use of Eq.~(\ref{tau/B}). For clarity some low temperature
        measurements are not shown.}
	\protect\label{tau(B)}
\end{figure}

The in-plane penetration depth $\lambda_{ab}(T)$ was
extracted from field-dependent measurements carried out at different
temperatures with the field applied at $\delta = 45^{\circ}$.
At this angle a reversible signal was observed over a large field range
down to 10 K, which allows the determination of $\lambda_{ab}(T)$ in a
wide temperature range.
The reversible part of the torque signal, $\tau/B_{a} \propto M$, recorded
on sample \Roman{one}$b$ (after the second annealing in $^{16}$O) at different
temperatures is shown as a function of $B_{a}$ in Fig.~\ref{tau(B)}.
The logarithmic field dependence,
characteristic for a type \Roman{two} superconductor, is clearly seen
for small applied fields. In this field regime the reversible torque
is given by \cite{FarrellPRL88,HoferPRB99}
  \begin{eqnarray}
   	\frac{\tau}{B_{a}} & = & \frac{\alpha V \Phi_{0}}{8 \pi^{2} \mu_{0}
   	\lambda_{ab}^{2}(T)} \left(1 - \frac{1}{\gamma^{2}} \right)
   	\frac{\sin2\delta}{\epsilon(\delta)}
   	\nonumber \\
   	& & \cdot \ln\left(\frac{\beta \xi_{ab}^{2}(T)
   	\epsilon(\delta)}{\Phi_{0}}B_{a}\right).
   	\label{tau/B}
  \end{eqnarray}
$\gamma = \sqrt{m_{c}^{\ast}/m_{ab}^{\ast}}$ is the effective mass anisotropy
, $\xi_{ab}(T)$ is the in-plane correlation length, and
$\epsilon(\delta) = (1/\gamma^{2} \sin^2\delta + \cos^2
\delta)^{1/2}$. The numerical factors $\alpha$ and $\beta$ depend on
the specific model \cite{FarrellPRL88,HoferPRB99}. Equation~(\ref{tau/B}) is
valid only for fields $B_{a} < B^{\ast}(T)$, where
the data points in Fig.~\ref{tau(B)} lie on a straight line. As an example
$B^{\ast}(T = 20.5\mbox{ K})$ is indicated by an arrow.
For $B_{a} > B^{\ast}(T)$ the condition $B_{a} \ll \Phi_{0}/[\xi_{ab}^{2}(T)
\epsilon(\delta)]$ for Eq.~(\ref{tau/B}) to be valid
\cite{KoganPRB88,HoferPRB99} is no longer fulfilled.

For $\delta = 45^{\circ}$ the dependence of $\tau/B_{a}$ in
Eq.~(\ref{tau/B}) on $\gamma$ is very weak for large $\gamma$
values, since $\epsilon(45^{\circ}) \simeq \cos(45^{\circ})$.
\noindent
Nevertheless, in order
to extract $\lambda_{ab}(T)$ from field-dependent measurements
by use of Eq.~(\ref{tau/B}), a determination of $\gamma$ is favorable.
Therefore, we performed angular-dependent torque measurements
close to $T_{c}$. Equation~(\ref{tau/B}) can also be used to analyze
angular-dependent torque data, provided that the measurements are performed
at $B_{a} \leq B^{\ast}(T)$.
In order to obtain a fully reversible signal over the whole angular regime
in these small fields, we applied an additional AC field perpendicular to
$\vec{B_{a}}$ in order to
enhance the relaxation processes \cite{WilleminPRB98}. From these
measurements $\gamma$ was determined for each sample. The penetration
depth
$\lambda_{ab}^{-2}(T)$ was then extracted from the slope of the linear
part of the field-dependent data (solid lines in Fig.~\ref{tau(B)}), using
Eq.~(\ref{tau/B}) with $\gamma$ fixed.

\begin{figure}[htb]
        \centering
        \includegraphics[width=0.8\linewidth]{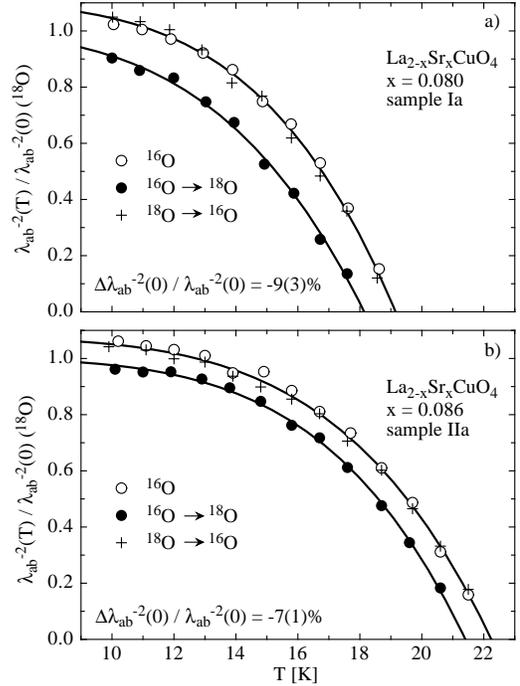}
        \caption[~]{Normalized in-plane penetration depth \newline
        $\lambda_{ab}^{-2}(T) / \lambda_{ab}^{-2}(0)(^{18}$O) for samples
        \Roman{one}$a$ and \Roman{two}$a$. $\lambda_{ab}^{-2}(0)$ is
        determined by extrapolating the data to $T
        = 0$ K, using the power law
        $\lambda_{ab}^{-2}(T) = \lambda_{ab}^{-2}(0)(1 - (T/T_{c})^{n})$
        (solid lines). The data of the back-exchanged sample demonstrate the
        reproducibility of the exchange procedure.}
	\protect\label{lambda-shift}
\end{figure}

Figure~\ref{lambda-shift} displays
$\lambda_{ab}^{-2}(T)$ for the samples
\Roman{one}$a$ and \Roman{two}$a$. The temperature dependence is well
described by the
\begin{table}[htb]
    \centering
	\caption[~]{Summary of the OIE results of the four
	La$_{2-\mbox{\footnotesize{x}}}$Sr$_{\mbox{\footnotesize{x}}}$CuO$_{4}$ 
	singl
e crystals with $x = 0.080$ (samples \Roman{one}$a$ and
    \Roman{one}$b$) and $x = 0.086$ (samples \Roman{two}$a$ and
\Roman{two}$b$).}
	\protect\label{Tab1}
    \begin{tabular}{lcccccc}
    	sample & mass & $T_{c}$ ($^{16}$O) & $T_{c}$ ($^{18}$O) &
    	$\frac{\Delta T_{c}}{T_{c}}$ &$\alpha_{O}$
    	&$\frac{\Delta \lambda_{ab}^{-2}(0)}{\lambda_{ab}^{-2}(0)}$  \\
            & [$\mu$g] & [K] & [K] & [\%] & & [\%]  \\
    	\hline
    	\Roman{one}$a$  & 9.6  &19.52(5) & 18.45(5)
    	& -5.5(4) & 0.45(3) & -9(3)  \\
    	\Roman{one}$b$  & 12.1 &19.68(5) & 18.50(5)
    	& -6.0(4) & 0.49(3) & -11(3) \\
    	\Roman{two}$a$  & 3.4  &22.40(5) & 21.26(5)
    	& -5.1(3) & 0.42(3) & -7(1)  \\
    	\Roman{two}$b$  & 3.8  &22.11(5) & 21.11(5)
    	& -4.5(3) & 0.37(3) & -10(1) \\
    	\hline
    	mean \Roman{one} &      &  & & -5.7(3) & 0.47(2) & -10(2) \\
    	mean \Roman{two} &      &  & & -4.8(2) & 0.40(2) & -8(1)  \\
    \end{tabular}
\end{table}
\noindent
power law
$\lambda_{ab}^{-2}(T) = \lambda_{ab}^{-2}(0)(1 - (T/T_{c})^{n})$ with
an exponent $n \approx 5$. The fact that $\lambda_{ab}(T)$ can be
determined
down to $T \simeq 0.5 \cdot T_{c}$ justifies the extrapolation of
$\lambda_{ab}(T)$ to $T = 0$ K using this empirical power law. By
normalizing the extracted $\lambda_{ab}^{-2}(T)$ values to the low
temperature values $\lambda_{ab}^{-2}(0)$ obtained for the $^{18}$O
exchanged samples, any uncertainties in determining the
sample volume $V$ are avoided. From Fig.~\ref{lambda-shift} it is evident
that not only
$T_{c}$ but also $\lambda_{ab}^{-2}(0)$ shifts upon replacing $^{16}$O by
$^{18}$O. The shifts are found to be
$\Delta \lambda_{ab}^{-2}(0) / \lambda_{ab}^{-2}(0) = -9(3)\%$ and
$-7(1)\%$ for the samples \Roman{one}$a$ ($x = 0.080$) and
\Roman{two}$a$ ($x = 0.086$), respectively. Again, the data obtained on the
back-exchanged samples (cross symbols) coincide with the data recorded after
the first $^{16}$O annealing. This demonstrates the reproducibility of the
exchange
procedure.
A summary of the isotope effects obtained for all four
samples is given in Table~\ref{Tab1}.

Since $\lambda_{ab}^{-2}(0) \propto n_{s}/m_{ab}^{\ast}$, the oxygen
isotope shift of the penetration depth is due to a shift of $n_{s}$
or $m_{ab}^{\ast}$
  \begin{equation}
   	\Delta \lambda_{ab}^{-2}(0) / \lambda_{ab}^{-2}(0)
   	= \Delta n_{s} / n_{s} - \Delta m_{ab}^{\ast} / m_{ab}^{\ast}.
   	\label{Deltalambda}
  \end{equation}
There are several independent experiments
\cite{ZhaoNAT97,ZhaoJPCM98,ZhaoPRB95} on
La$_{2-\mbox{\footnotesize{x}}}$Sr$_{\mbox{\footnotesize{x}}}$CuO$_{4}$
samples which have shown that the change of $n_{s}$ during the exchange
procedure is negligible. From the present study, further evidence
that $n_{s}$ is unchanged during the isotope exchange is given by the
complete reproducibility of the exchange procedure.
It is hardly possible that $n_{s}$ changes upon $^{18}$O substitution,
but adopts again exactly the same value after
the back-exchange as in the $^{16}$O annealed sample.
We thus conclude, that any change in $n_{s}$ during
the exchange procedure is negligible, and that the change of the in-plane
penetration depth is mainly due to the isotope effect on the in-plane
effective
mass $m_{ab}^{\ast}$.

The observed OIE on $m_{ab}^{\ast}$ gives strong evidence
that lattice effects play an important role in high-$T_{c}$
superconductivity. A possible explanation for the strong dependence
of $m_{ab}^{\ast}$ on the oxygen isotope mass can be given by
a model of small bipolarons, where $m_{ab}^{\ast} \propto m_{ab} \exp(g^{2})$
($m_{ab}$ is the bare hole mass) \cite{AlexandrovBOOK95}. Since the polaronic
enhancement factor $g^{2} \propto 1/\omega$ depends on the characteristical
optical phonon frequency $\omega$ \cite{AlexandrovBOOK95}, a change
of the frequency leads to a change of $m_{ab}^{\ast}$.
The exponent of the total (copper and oxygen) isotope effect on
$m_{ab}^{\ast}$, $\beta_{\mbox{\footnotesize{tot}}} = \beta_{Cu} +
\beta_{O}$, is then given by
  \begin{equation}
   	\beta_{\mbox{\footnotesize{tot}}}
     	  = - (\Delta m_{ab}^{\ast} / m_{ab}^{\ast}) /
   	        (\Delta
M_{\mbox{\footnotesize{r}}}/M_{\mbox{\footnotesize{r}}})
   	      = - 0.5 g^{2}.
   	\label{beta}
  \end{equation}
The effective reduced mass $M_{\mbox{\footnotesize{r}}}$ is a
complicated function of
$M_{O}$ and $M_{Cu}$, depending on the symmetry of the modes. From the
experimentally
observed shifts in $\lambda_{ab}^{-2}(0)$ we can determine the exponent
$\beta_{O} = - (\Delta m_{ab}^{\ast} / m_{ab}^{\ast}) /
(\Delta M_{O} / M_{O})$. Taking a mean value
of $\Delta \lambda_{ab}^{-2}(0) / \lambda_{ab}^{-2}(0) \simeq -9\%$ (see
Table~\ref{Tab1}) and using  Eq.~(\ref{Deltalambda}), we find
$\beta_{O} \simeq (\Delta \lambda_{ab}^{-2}(0) / \lambda_{ab}^{-2}(0))
 / (\Delta M_{O} / M_{O}) \simeq - 0.7$. A universal relation
between $T_{c}$ and $\lambda_{ab}^{-2}(0)$ was experimentally
found in the cuprates, showing $T_{c} \propto \lambda_{ab}^{-2}(0)$ in the
deeply underdoped regime \cite{SchneiderIJMP93}. If we consider a
slightly weaker dependence of $T_{c}$ on $\lambda_{ab}^{-2}(0)$ for the
doping range investigated, we can assume $T_{c} \propto
[\lambda_{ab}^{-2}(0)]^{t}$ with $t < 1$. We thus find $\alpha_{O}
\simeq -t \beta_{O}$ (with $t \simeq 0.6$ from our experiment) and
$\alpha_{Cu} \simeq -t \beta_{Cu}$. Since $\alpha_{Cu}$ was found to be
similar to
$\alpha_{O}$ \cite{FranckPRL93,ZhaoPRB96}, it is plausible to assume
that $\beta_{Cu} \simeq \beta_{O}$ as well. We then find
$\beta_{\mbox{\footnotesize{tot}}} \simeq 2 \beta_{O} \simeq -1.4$, and
thus  $g^{2} \simeq 2.8$ from
Eq.~(\ref{beta}).
On the other hand, $g^{2}$ can also be determined from optical
conductivity data, which according to the small polaron model
show a maximum at $E_{m} = 2g^{2}\hbar \omega$ \cite{AlexandrovBOOK95}. In
La$_{2-\mbox{\footnotesize{x}}}$Sr$_{\mbox{\footnotesize{x}}}$CuO$_{4}$
this energy was found to be $E_{m} = 0.44$ eV for $x = 0.06$ and
$E_{m} = 0.24$ eV for $x = 0.10$ \cite{BiPRL93}. For our samples
with $x$ lying between these two values, we expect $E_{m} \simeq 0.34$
eV. With
$\hbar \omega \simeq 0.06$ eV \cite{ZhaoJPCM98} we thus find $g^{2}
\simeq 2.8$, in agreement with the
magnitude of $g^{2}$ deduced from the OIE on $m_{ab}^{\ast}$.

In summary we have studied the OIE on $T_{c}$ and
on $\lambda_{ab}^{-2}(0)$ in underdoped
La$_{2-\mbox{\footnotesize{x}}}$Sr$_{\mbox{\footnotesize{x}}}$CuO$_{4}$
microcrystals using a highly sensitive torque magnetometer. The
reproducibility of the isotope exchange procedure, as checked by
back-exchange, gives evidence for a complete
isotope exchange in the single crystals. The isotope
shift in $\lambda_{ab}^{-2}(0)$ is attributed to a shift in the
in-plane effective mass $m_{ab}^{\ast}$. For $x = 0.080$ and $x =
0.086$ we find $\Delta m_{ab}^{\ast} / m_{ab}^{\ast} = - 10(2)\%$
and $-8(1)\%$, respectively.
The OIE on $m_{ab}^{\ast}$
gives strong evidence that lattice vibrations play an important role
in the occurrence of high temperature superconductivity.

We are grateful to A. Revcolevschi for providing the large single
crystals on which the preliminary studies were performed. Fruitful
discussions with C. Rossel are acknowledged. This work was
partly supported by SNSF (Switzerland), NEDO and CREST/JST (Japan).
One of the authors (T.S.) would like to thank JSPS for financial support.


\begin{thebibliography}{10}

\bibitem{BednorzZPBCM86}
J.G.~Bednorz and K.A.~M\"{u}ller, Z. Phys. B {\bf64}, 189 (1986).

\bibitem{FranckBOOK94}
For a review see J.P.~Franck, in: \emph{Physical Properties of High
Temperature Superconductors IV}, ed. D.M.~Ginsberg (World
Scientific, Singapore, 1994) (p189-293).

\bibitem{BatloggPRL87}
B.~Batlogg \emph{et al.}, Phys. Rev. Lett. {\bf58}, 2333 (1987).

\bibitem{BatloggPRL287}
B.~Batlogg \emph{et al.}, Phys. Rev. Lett. {\bf59}, 912 (1987).

\bibitem{CrawfordPRB90}
M.K.~Crawford \emph{et al.}, Phys. Rev. B {\bf41}, 282 (1990).

\bibitem{ZechNAT94}
D.~Zech \emph{et al.}, Nature (London) {\bf371}, 681 (1994).

\bibitem{FranckPRL93}
J.P.~Franck, S.~Harker, and J.H.~Brewer, Phys. Rev. Lett. {\bf71},
283 (1993).

\bibitem{ZhaoPRB96}
G.M.~Zhao \emph{et al.}, Phys. Rev. B {\bf54}, 14 956 (1996).

\bibitem{ZhaoNAT97}
G.M.~Zhao \emph{et al.}, Nature (London) {\bf385}, 236 (1997).

\bibitem{MartinPC95}
A.A~Martin and M.J.G.~Lee, Physica (Amsterdam) {\bf254C}, 222 (1995).

\bibitem{LiPRB93}
Q.~Li \emph{et al.}, Phys. Rev. B {\bf47}, 2854 (1993).

\bibitem{WilleminJAP98}
M.~Willemin \emph{et al.}, J. Appl. Phys. {\bf83}, 1163 (1998).

\bibitem{FarrellPRL88}
D.E.~Farrell \emph{et al.}, Phys. Rev. Lett. {\bf61}, 2805 (1988).

\bibitem{HoferPC98}
J. Hofer \emph{et al.}, Physica (Amsterdam) {\bf297C}, 103 (1998).

\bibitem{KoganPRB88}
V.G.~Kogan, M.M.~Fang, and S.~Mitra, Phys. Rev. B {\bf38}, R11 958
(1988).

\bibitem{BlatterRMP94}
G.~Blatter \emph{et al.}, Rev. Mod. Phys. {\bf66}, 1125 (1994).

\bibitem{SasagawaPRL98}
T.~Sasagawa \emph{et al.}, Phys. Rev. Lett. {\bf80}, 4297 (1998).

\bibitem{ZhaoJPCM98}
G.M.~Zhao \emph{et al.}, J. Phys.: Condens. Matter {\bf10}, 9055 (1998).

\bibitem{HoferPRB99}
T.~Schneider \emph{et al.}, Eur. Phys. J. B {\bf3}, 413 (1998);
J.~Hofer \emph{et al.}, Phys. Rev. B {\bf60}, 1332 (1999).

\bibitem{WilleminPRB98}
M.~Willemin \emph{et al.}, Phys. Rev. B {\bf58}, R5940 (1998).

\bibitem{ZhaoPRB95}
G.M.~Zhao \emph{et al.}, Phys. Rev. B {\bf52}, 6840 (1995).

\bibitem{AlexandrovBOOK95}
A.S.~Alexandrov and N.F.~Mott, \emph{Polarons and Bipolarons} (World
Scientific, Singapore, 1995).

\bibitem{SchneiderIJMP93}
T.~Schneider and H.~Keller, Int. J. Mod. Phys. {\bf8}, 487 (1993).

\bibitem{BiPRL93}
X.X.~Bi and P.C.~Eklund, Phys. Rev. Lett. {\bf70}, 2625 (1993).

\end{thebibliography}
\end{document}